\documentclass[a4paper, 10pt, fleqn, twocolumn]{article}

\usepackage{amsmath}
\usepackage{graphicx,epsf}
\usepackage{hyperref}
\usepackage[superscript]{cite}
\usepackage{color}

\newcommand{\rev}[1]{{\color{black}#1}}
\newcommand{\ds}{\displaystyle}

\usepackage[switch, modulo]{lineno}

\begin{document}

\title{Photogalvanic and photon drag phenomena in superconductors and hybrid superconducting systems}


\author{S.~V.\ Mironov, A.~I.\ Buzdin, O.~B.\ Zuev, M.~V.\ Kovalenko and A.~S.\ Mel'nikov}

\maketitle

\begin{abstract}
In this paper we review the recent progress in theoretical understanding of the peculiarities of  photogalvanic phenomena, photon drag and inverse Faraday effects in superconductors and hybrid superconducting structures. Our study is based on the time-dependent Ginzburg-Landau (TDGL) theory with a complex relaxation constant which provides the simplest description of the mechanisms of the second-order nonlinear effects in the electrodynamic response and related mechanisms of generation of dc photocurrents, magnetic moment and switching between different current states under the influence of electromagnetic radiation of various polarization.
\end{abstract}

\section*{Introduction}

Different aspects of nonlinear electrodynamics of superconductors attracted the interest of both experimentalists and theoreticians for many decades. These studies revealed a number of fascinating physical phenomena and suggested various directions for applications. Most of these phenomena arise from the obvious fact that the superfluid density is partially suppressed by the supercurrents induced by the electromagnetic wave and the modulation of this quantity results in the nonlinear corrections to the microwave response which are odd in the amplitude of the electromagnetic signal. The simplest example is the third order nonlinearity which is responsible for the generation of the third harmonic in the supercurrent which can be obtained already within the Ginzburg-Landau formalism. 
Much less is known about the electrodynamic response of the superconducting condensates containing the even order terms in the wave amplitude. Meanwhile such nonlinearities can be not only exceptionally interesting from the fundamental point of view but provide very promising route to a number of applications in superconducting optoelectronics or optofluxonics. The goal of our recent works 
\cite{IFE1,plast,croit,croit2,IFE2,Photon_drag,mironov-diode,Oleg}
has been to fill this gap and suggest a theoretical description of possible nonlinear effects originating from the second order nonlinear response in superconducting materials and hybrid structures. 

In general, the even order nonlinearities in the electrodynamic response can give rise to two experimentally measurable effects\cite{Belinicher_rect,Ivchenko_rect,Glazov_rect,tokura,taras,perel,barlow,normantas,taras2,gurevich1,gurevich2,ivchenko,taras3}: (i) generation of the even harmonics of the primary wave frequency, (ii) generation of rectified dc current. At this point it is useful to note an important difference between the photogalvanic phenomena in normal and superconducting systems related to the dissipationless nature of supercurrents. Indeed, the dc  photocurrents generated by the electromagnetic pulse in nonsuperconducting material should decay at a certain characteristic time after switching off the pulse due to the Ohmic losses. In superconducting systems these currents in principle can persist even without supporting primary electromagnetic wave which in this case plays the role of trigger switching the superconductor between different current carrying states. Certainly, such switching can occur only at appropriate conditions and the resulting stable current states must be protected from the relaxation processes. An obvious way to realize this protection is to switch between the superfluid states which differ by their topology. A well known example of such topologically protected state in superfluid systems is the state with nonzero vorticity or winding number. In other words, the photoinduced currents in superconductors can be in principle used to create and switch the vortex configurations \cite{plast,yeh1,yeh2}.  This possibility looks very tempting as it provides an alternative to the switching by magnetic field being at the same time more fast and, thus, more perspective for any logic based on manipulation of fluxons (such as RSFQ logic \cite{Likharev_Semenov}).    

Hereafter we focus on the simplest lowest order nonlinearity of the material namely on the current contribution of the second order in the electric field of the incident electromagnetic wave.   In complex valued Fourier components a general expression describing the linear and second order nonlinear response in  a certain conducting medium takes the form:
\begin{equation}
j_n = \sigma_{nm}E_m e^{i(\mathbf{k}\mathbf{r}-\omega t)}+ \zeta_{nml}E_m E^*_l + \tilde\zeta_{nml}E_m E_l e^{2i(\mathbf{k}\mathbf{r}-\omega t)} \ ,
\end{equation}  
where $j_n$, $E_{n}$ with $n=x,y,z$ are current density and electric field components, $\sigma_{nm}$ is the conductivity tensor, $\zeta_{nml}$ and $\tilde\zeta_{nml}$ are the third rank tensors describing dc and second harmonic contributions in the nonlinear response and the repeated indices assume the summation. 
Considering the mechanisms responsible for the nonzero tensor $\zeta_{nml}$ we can first mention the media with a certain polar vector $\mathbf{d}$ breaking the material isotropy. In this case the tensor can be written as follows:
\begin{equation}
 \zeta_{nml}= C_1 d_n\delta_{ml}+ C_2 \delta_{nm} d_l \ ,
\end{equation}    
where $\delta_{ml}$ is the Kronecker symbol and $C_1$, $C_2$ are material dependent constants.
We get here an example of the material with the intrinsic diode effect or nonreciprocal electromagnetic response. 
In nonsuperconducting materials these phenomena are well known to reveal, e.g., in the linear dependence of resistivity 
on the applied electric field or current density and, thus, obvious asymmetry of the current-voltage characteristics.
Generalizing the above expressions for the case of superconductors we can expect that the corresponding superconducting diode effect should result in the asymmetry of the nonlinear relation between the current density and the superfluid velocity and also in the directional dependence of the critical depairing current destroying the superconducting state. 
The manifestations of the diode effect in different superconducting hybrid systems has been discussed for several decades for Josephson junctions \cite{Krasnov_DE}, ratchet-type pinning \cite{Villegas_DE} etc. and recently revisited in connection with the experiments in multilayers \cite{ando} (see also \cite{Nadeem_rev} for review). One of the popular scenario of the intrinsic diode effect relates to the so-called magnetochiral anisotropy arising in the presence of both spin-orbit and Zeeman-type effects in low dimensional proximitized systems. In this case the above polar vector
takes the form of a vector product of the Zeeman (or exchange) field and the vector perpendicular to the interface of materials in the layered structure originating from the Rashba-type expression for the interface spin-orbit interaction \cite{daido,he,edel1,edel2,persh,mironov-B}.
Another exemplary mechanism responsible for the appearance of the second order nonlinear response can originate from the   
wave vector $\mathbf{k}$ of the electromagnetic radiation and the appropriate tensor takes the form:
$\zeta_{nml}= W_{nmls}k_s$. Such expressions are typical for the systems revealing the so called photon drag effect caused by the transfer of momentum from the electromagnetic radiation to the electrons. Circular polarization of light provides one more interesting version of the second order nonlinear phenomenon associated with the transfer of the angular momentum and resulting inverse Faraday effect predicted first by L.~P.~Pitaevsky \cite{Pitaevskii61}. The incident electromagnetic wave in this case generates the magnetic moment determined by the helicity of radiation.

We now briefly discuss one important peculiarity of the superconducting systems which affects strongly the physics of the electromagnetic response and in particular the nonlinear effects. The point is that in a superconducting material we get two different types of carriers responsible for the flowing charge current: Cooper pairs and quasiparticles. Both these types of charged fluids are affected by the applied electric field and also by the gradients of their chemical potentials. For a normal metal with single type of charge carriers the presence of the gradient of chemical potential generates the diffusion current which  can be simply summed up with the one generating by the electric field and finally we obtain the total electrochemical field moving the electrons. It is this field which enters in fact the above expressions for the current in normal metal. In a system with two types of carriers the chemical potentials in nonequilibrium transport state can be different and we can no more introduce a single electrochemical field acting on the carriers. This is why in superconductors it is convenient to introduce two sources generating the current: (i) electrochemical field accounting for the gradient of chemical potential of quasiparticles and (ii) gradient of difference in the chemical potential $\tilde \Phi$ of quasiparticles and Cooper pairs. The latter quantity is also known as the charge imbalance potential and appears to be nonzero in many problems of nonequilibrium  superconductivity.  
Thus, considering nonlinear effects we should now take account of additional mechanism of nonlinearity and rewrite the expressions for the current including both the fields $\mathbf{E}$ and $\nabla \tilde\Phi$. 

Another important peculiarity of superconductors is that the dc photocurrents generated according to any of the above mechanisms can be screened by the supercurrents flowing at zero frequency. These currents can flow in the opposite directions and contain two contributions: the one from the vector potential which gives just the Meissner screening and the one from the gradient of superconducting phase which is needed to guarantee the current continuity condition. 

Note also that all photogalvanic and photon drag phenomena discussed above preserve at temperatures slightly above the critical temperature due to the presence of superconducting fluctuations. Some examples of microscopic and phenomenological models treating this situation can be found in \cite{boev1,boev2,boev3,Sonowal,Parafilo_2025}.

After these brief introductory notes we proceed with the discussion of several particular scenarios illustrating generation of the dc supercurrents by the electromagnetic wave.

\section*{Photogalvanic phenomena in superconducting hybrids with the diode effect}

A universal description of the intrinsic diode phenomena in superconducting state can be obtained within the modification of the phenomenological Ginzburg-Landau (GL) theory. The presence of the preferable current direction can be accounted in this theory
by introducing the terms odd in the spatial derivatives of the order parameter into the GL functional. To specify the form of these terms it is convenient to keep in mind some particular scenario of the nonreciprocal effects. As such scenario we choose here the bilayer structure consisting of a thin superconducting film covered by a certain magnetic material. The latter is assumed to be responsible for an exchange field acting on the spins of the superconducting electrons. This spin splitting field can be in principle momentum dependent either due to the spin orbit effects at the sample interface or due to the nontrivial spatial symmetry of the orbitals with ordered localized spins in the magnetic layer (which appears, e.g., in altermagnetic materials).
To guess the form of the gradient terms in the effective GL functional averaged over the bilayer thickness it is instructive to start from the averaged single-particle Hamiltonian omitting first the attraction between electrons:
\begin{equation}\label{Eq3}
\begin{array}{c}{\ds 
\hat H = \frac{\hat{\mathbf{p}}^2}{2m_e} + \hat{\sigma}_n h_n +\alpha_{nm} \hat{\sigma}_n \hat{p}_m ~~~~~~~~~~~~~~~~~~~~~~~~~~}\\{\ds~~~~~~~~~~~~~~~~~~~+\gamma_{nmsl}\hat{\sigma}_{n} \hat{p}_{m} \hat{p}_{s} \hat{p}_{l} +\tilde h_{nms}\hat{\sigma}_{n} \hat{p}_{m} \hat{p}_{s} \ ,}
\end{array}
\end{equation}
where {$\hat{\bf p}$ and $m_e$ are the electron momentum and mass, respectively,} $\hat{\sigma}_{n}$ with ${n}=x,y,z$ are the Pauli matrices and the summation is assumed over the repeated indices. The field ${\bf h}$ is the standard exchange field, {the tensor} $\alpha_{nm}$ describes the terms linear in the electron momentum and, thus, can correspond, e.g., to the Rashba spin-orbit coupling at the interface (for $\alpha_{nm}=\alpha n_{0s}\epsilon_{snm}$ with Levi-Civita tensor $\epsilon_{snm}$ and the {unit} vector {${\bf n}_0$} normal to the interface), {the tensors} $\gamma_{nmsl}$ and $\tilde h_{nms}$ describe the momentum dependence of the exchange field mentioned above. The GL free energy for a singlet pairing can not depend of course on the spin operator and constructing the GL functional one should replace the electronic spin in the above Hamiltonian by the averaged spin polarization induced by one of the spin splitting fields. Thus, the recipe to get the relevant GL gradient terms should look as follows: we just take all possible products of different pairs of  spin splitting fields in the above Hamiltonian and replace the momentum $\hat{\bf p}$ by the gradient of the GL order parameter elongated by the vector potential to guarantee the gauge invariance. Finally, we obtain the following possible contributions to the odd gradient part of the free energy density:
\begin{equation}\label{Eq4}
F_1\sim f_1\Psi^*\alpha_{nm} h_{n} \hat{D}_{m} (1+{\eta} \hat{\bf D}^2) \Psi \ ,
\end{equation}
\begin{equation}\label{Eq5}
F_2\sim f_2\Psi^*\gamma_{nmsl} h_{n} \hat{D}_{m} \hat{D}_{s} \hat{D}_{l}\Psi \ ,
\end{equation}
\begin{equation}\label{Eq6}
F_3\sim f_3\Psi^*\alpha_{nm} \tilde h_{nsl} \hat{D}_{m} \hat{D}_{s} \hat{D}_{l} \Psi \ ,
\end{equation}
where $\hat{\bf D} = -i\hbar\nabla - 2e\mathbf{A}{/c}$ {is the gauge-invarient momentum operator} ({here} $e < 0$). The term $F_1$ describes the joint effect of the Rashba spin-orbit term and exchange field, $F_2$ describes the structure with higher order in momentum spin-orbit interaction and exchange field, and $F_3$ corresponds to the Rashba-type coupling in the bilayer with d-wave altermagnetic material.
\rev{
The expressions {(\ref{Eq4})-(\ref{Eq6})} have been constructed only from the symmetry arguments, namely from the fact that all contributions to the free energy should be scalars and, thus, the related gradient terms should be proportional to the scalar products of different momentum dependent exchange field terms in Eq.~(\ref{Eq3}). This line of symmetry based reasoning can not exclude, of course, the dependence of the prefactors $f_1$, $f_2$, $f_3$ in expressions {(\ref{Eq4})-(\ref{Eq6})} on the convolutions of the tensors entering the Eq.~(\ref{Eq3}). The full expressions for the coefficients in {(\ref{Eq4})-(\ref{Eq6})} can be obtained only on the basis of microscopic theory. 
To illustrate possible nonlinear dependence of the prefactors on the tensor convolutions one can consider, e.g., such direct derivation of the invariant $F_1$ presented in \cite{Plastovets2}. For this purpose, we should restrict ourselves by the first three terms in the Eq.~(\ref{Eq3}) and exploit a standard Gor'kov mean-field approach which gives us the self-consistency equation for superconducting gap:
\begin{equation} \label{Eq2}
\Delta^*({\bf r}) = g  T\sum_{\omega_n} \text{Tr} \left[\hat{\sigma}_+ \hat{F}^\dagger({\bf r},{\bf r})\right],
\end{equation}
where $g$ is the BCS coupling constant, {$T$ is the system temperature,} $\hat{\sigma}_+=(\hat{\sigma}_x+i\hat{\sigma}_y)/2$ is the combination of Pauli matrices in the spin space, and $\hat{F}^\dagger$ is the anomalous Green function which should be obtained from the Gor'kov equations (see \cite{Plastovets2}). 
To get the gradient terms in the Ginzburg-Landau equation 
it is sufficient to expand the Eq. (\ref{Eq2}) up to linear in $\Delta$ terms: 
 $L^{-1}({\bf q})\Delta^*({\bf q})=0$, where the propagator reads as 
\begin{equation}\label{L}
L^{-1}=\frac1g+T\sum\int\limits_{\omega_n, {\bf p}} \text{Tr}  \left[
\hat{\sigma}_+ \hat{\bar{G}}_{0}\left({\bf p}-\frac{{\bf q}}{2}\right)i\hat{\sigma}_2\hat{{G}}_0\left({\bf p}+\frac{{\bf q}}{2}\right)\right].
\end{equation}
Choosing the particular configuration of the quantization axis as ${\bf h}||{\bf x}_0$ and ${\bf n_{0}}||{\bf z}_0$ we get  
\begin{equation}
\hat{G}_0({\bf p}) = \frac{1}{D_0}
\begin{pmatrix}
    i\omega_n-\xi_p & h-\alpha (p_y+ip_x) \\
    h-\alpha (p_y-ip_x) & i\omega_n-\xi_p
\end{pmatrix}, 
\end{equation}
where ${D_0=(i\omega_n-\xi_p)^2-h^2-\alpha^2 p^2+2\alpha h p_y}$; the normal metal spectrum is ${\xi_p}=p^2/2m_e-\mu$; {$\mu$ is the chemical potential;} and ${\bf p}=(p_x,p_y)$ {is the momentum vector}. To restore the GL free energy functional we should expand Eq.~(\ref{L}) in powers of momentum $q$ and multiply 
 $L^{-1}$ by $|\Delta({\bf q})|^2$.
 For weak Zeeman field, namely $h~< \alpha p_F\ll T_{c0}$ we find
\begin{equation} \label{GL_q}
F_\text{S}=
\\ 
\Big(a_0+ a_{h} {\bf h}^2 + {\bf q} [{\bf n}_{0}\times{\bf h}_{||}] \big(a_1+a_3 {\bf q}^2\big)+a_2{\bf q}^2  + a_4{\bf q}^4\Big) \Delta^2,
\end{equation}
where ${\bf h}_{||}$ is the in-plane component of the Zeeman field{,} the coefficients {read}
\begin{eqnarray} \notag
 a_0={N(0)} \ln\left(\frac{T}{T_{c0}} \right); ~ a_2 =\frac{7\zeta(3)v_F^2{N(0)}}{32\pi^2T_{c0}^2}; \\  \notag a_4 = \frac{-93\zeta(5)v_F^4{N(0)}}{2048\pi^4 T_{c0}^4}; \quad a_1 = \alpha C_\alpha \frac{ \zeta(5) {N(0)} }{32\pi^4 T_{c0}^2}; \\ \notag  a_3 =-   \alpha C_\alpha  \frac{1905\zeta(7) v_F^2{N(0)}}{2048\pi^6 T_{c0}^4}; \quad\quad\quad\quad \\ \label{a_i}
    a_{h} = {N(0)} \left(\frac{7\zeta(3)}{4\pi^2 T^2_{c0}} - \frac{31\zeta(5)}{32\pi^4}\frac{\alpha^2 p_F^2}{T^4_{c0}} \right)+\mathcal{O}\left(\frac{h^2}{T^2_{c0}}\right),
\end{eqnarray}
{the value $C_\alpha$ is defined as} $C_\alpha=\alpha^2p_F^2/T_{c0}^2$ {and $N(0)$ is the density of states at the Fermi level}. Replacing now ${\bf q}$ by {the operator} $\hat{\mathbf{D}}$ and keeping in mind that $\Delta\propto\Psi$ one can easily restore the form of the gradient terms in Eq.~(\ref{Eq4}). Clearly, the above derivation shows that the coefficient $f_1$ quadratically depends on $\alpha$ in the considered model and parameter range. Note that similar results have been previously obtained in \cite{edel1, edel2, Levchenko}. Interestingly, in the strong SOC regime the dependence of the coefficient $f_1$ on $\alpha$ disappears \cite{edel1, Levchenko}.


 Using now the London-type model, i.e. neglecting the changes in the order parameter absolute value, and introducing the superconducting phase $\chi$ and the superfluid velocity $v = (\hbar/2m)(\nabla\chi - 2e\mathbf{A}/\hbar c)$ we find the analogue of the above nonlinear expression for the current:
\begin{equation}
j_n = Q_{nm}v_{m} e^{i(\mathbf{k}\mathbf{r}-\omega t)}+ R_{nml}v_{m} v^*_{l} + \tilde R_{nml}v_{m} v_{l} e^{2i(\mathbf{k}\mathbf{r}-\omega t)} \ .
\end{equation} 
Note that {here} we shifted the superfluid velocity (in other words the phase gradient) to exclude an additional linear in $v$ term which appears in the expression for {the value} $F_1$ so that the free energy minimum corresponds to the zero velocity $v$.

To apply this expression for the analysis of nonlinear electrodynamics and related nonreciprocal phenomena
we can consider a superconducting film of the thickness $d_s$ 
 irradiated by the linearly polarized electromagnetic wave with the wave vector perpendicular to the film surface. The thickness $d_s$ is assumed to be much larger than the interatomic distance to ensure the full electromagnetic wave reflection but, at the same time, much smaller than the London penetration depth. The latter condition allows to neglect the spatial distribution of the optically-induced electric current across the film. For  further calculations it is convenient to consider the magnetic field of the incident wave in the plane of the film in the form ${\bf B}={\rm Re}\left({\bf B}_\omega e^{-i\omega t}\right)$ where ${\bf B}_\omega$ is the complex amplitude of the wave and $\omega$ is the wave frequency. Then integrating the Maxwell equation for ${\rm curl}~{\bf B}$ over the film thickness we get:
\begin{equation}\label{Curr_omega}
2\left({\bf n}_{0}\times {\bf B_\omega}\right)=\frac{4\pi}{c}{\bf j}_{\omega}d_s,
\end{equation}
where ${\bf j}_{\omega}$ is the complex amplitude of the supercurrent at the frequency $\omega$ and the factor $2$ in the l.h.s. accounts the doubling of the amplitude of the magnetic field at the sample boundary due to the full reflection of the incident wave.
Considering the problem perturbatively we can express the superfluid velocity amplitude at the frequency of the incident wave from the relation $j_{n\omega} = Q_{nm}v_{m\omega}$ and substitute it into the expression for the rectified current
$j_{n,dc}= R_{nml}v_{m\omega} v^*_{l\omega}$.
 Further solution 
strongly depends on the proposed experimental setup and resulting boundary conditions. Indeed, if the irradiated sample is not included into the closed superconducting loop which would allow to get a circulating nonzero current  the dc photocurrent (given by the second term in the above expression) can not flow through the sample edges. The continuity of the current in this case requires the generation of the dc phase gradient which would compensate the dc photocurrent and as a result we obtain a nonzero phase difference at the edges of the sample. Thus, we get the superconducting phase battery \cite{samokh,aronov}. Assuming, e.g., for simplicity the tensor $Q_{nm}$ to be diagonal we find the resulting phase difference in the form: 
$\delta\chi\sim R_{nml}v_{m\omega} v^*_{l\omega} \sim R_{nml}j_{m\omega} j^*_{l\omega}$.
}

If, in opposite, the sample is included into the closed superconducting contour the dc photocurrent generates the circulating supercurrent and the flux in this loop
and changing the amplitude of the electromagnetic wave we can observe vortex entry/exit (see \cite{mironov-diode} for details).

\section*{Charge imbalance potential as the source of photogalvanic and drag phenomena}

Let us now switch to another mechanism responsible for the second order nonlinear response in superconductor mentioned in introduction, i.e. to the effects caused by the generation of the charge imbalance potential. The kernel $Q$ in the relation between the supercurrent and superfluid velocity should depend on the chemical potential $\mu$. Expanding this kernel up to the linear correction to $\mu$ we find:
\begin{equation}
\mathbf{j} =Q(\mu) \mathbf{v} +\frac{\partial Q}{\partial \mu} e\tilde\Phi \mathbf{v}.
\end{equation}
The second term here obviously gives us the desired nonlinearity since the potential $\tilde\Phi$ is also generated by the electromagnetic wave. Note that we omit here possible contribution to the photon drag phenomenon associated with the small changes in the total electronic density due to the electric field effect \cite{Radkevich}.

More careful calculations can be carried out within the time dependent GL theory. The dynamics of the superconducting order parameter $\Psi$ at temperatures $T$ near the superconducting transition critical temperature $T_c$ is described by the equation
\begin{equation}\label{GL_general}
\frac{\pi\alpha_{0}}{8}{\left(1+i\nu\right)} \left(\hbar\frac{\partial\Psi}{\partial t}+ 2ie\phi\Psi\right)=-\frac{\delta F}{\delta \Psi^*}, 
\end{equation}
where $F$ is the GL free energy 
\begin{equation}\label{F_general}
F=F_N+\int\left[-\alpha_{0} T_c\varepsilon\left|\Psi\right|^2+\frac{1}{4m_{0}}\left|{\bf D}\Psi\right|^2+\beta_{0}\left|\Psi\right|^4\right]d^3{\bf r}.
\end{equation}
Here $F_N$ is the system free energy in the normal state, $\alpha_{0}$ and $\beta_{0}$ are the standard GL coefficients, $m_{0} = \hbar^2/(4 \alpha_{0} T_c \xi_0^2)$, $\xi_0$ is the superconducting zero-temperature coherence length, $\varepsilon=1-T/T_c$, $\phi$ is the electrochemical potential of quasiparticles. The key ingredient of Eq.~(\ref{GL_general}) is the imaginary part {$\nu$} of the GL relaxation constant which arises due to the small electron-hole asymmetry and is responsible for the photon drag phenomena \cite{IFE1,plast,croit,IFE2,Photon_drag,mironov-diode,Oleg,plast2}. To make the physics beyond these phenomena more transparent we explicitly introduce the gap function $\Delta$ and the superconducting phase $\chi$ so that $\Psi=\Delta e^{i\chi}$. Then the superconducting current ${\bf j}_s$ reads
\begin{equation}\label{js_def}
{\bf j}_s=2e\Delta^2{\bf v},
\end{equation}
where ${\bf v}$ is the superfluid velocity defined above. The dynamics of $\Delta$ is controlled by the real part of the GL equation (\ref{GL_general}):
\begin{equation}\label{GL_real}
\begin{array}{c}{\ds
\frac{\pi\alpha_{0}}{8}\hbar\frac{\partial\Delta}{\partial t}-\left(\alpha_{0} T_c\varepsilon-m_{0}{\bf v}^2\right)\Delta }\\{}\\{\ds \quad\quad\quad\quad -\frac{\hbar^2}{4m_{0}}\nabla^2\Delta+\beta_{0}\Delta^3={\frac{\pi\alpha_0}{4}\nu} e \tilde\Phi \Delta,}
\end{array}
\end{equation}
where the charge imbalance potential $\tilde\Phi=\phi-\mu_p$ is the difference between the chemical potential of quasiparticles $\phi$ and the chemical potential of Cooper pairs $\mu_p=-\left(\hbar/2e\right)\partial\chi/\partial t$. In the absence of the electron-hole asymmetry (i.e. if one assumes ${\nu}=0$) the superconducting
current contains only the contributions which are odd with respect to the superconducting velocity $v$. The even contributions may arise only provided (i) ${\nu}\neq 0$ and (ii) the electromagnetic field of the incident wave induces the oscillations of the potential $\tilde\Phi$ which, according to Eq.~(\ref{GL_real}), then become transformed into the oscillations of the gap function $\Delta$. Note that the second condition becomes fulfilled only if the electric field ${\bf E}$ of the wave has a component perpendicular to the sample surface. To show this, we derive the equation for $\tilde\Phi$ which directly follows from the imaginary part of the GL equation (\ref{GL_general}) and the Maxwell equation ${\rm curl}~{\bf B}=\left(4\pi/c\right){\bf j}+\left(1/c\right)\partial {\bf E}/\partial t$, where ${\bf B}={\rm curl}~{\bf A}$ is the magnetic field and ${\bf j}$ is the total current containing both normal and superconducting components: ${\bf j}=\sigma {\bf E}+{\bf j}_s$ (here $\sigma$ is the normal state conductivity of the superconductor). Restricting ourselves to the contributions of the first order over the small parameter ${\nu}$ in the final expressions for the current, we neglect the deviation of the gap function
from the equilibrium value $\Delta_0=\sqrt{\left(\alpha_{0} T_c/\beta_{0}\right)\varepsilon}$ when dealing with the equation for $\tilde \Phi$ and, therefore, assume the London penetration depth $\lambda=\sqrt{m_{0}c^2/\left(8\pi e^2\Delta_0^2\right)}$ to be constant. Considering $e^{-i\omega t}$ processes and taking into account that ${\bf E}=-\nabla \phi-\left(1/c\right)\partial {\bf A}/\partial t$ we rewrite the expression for the supercurrent (\ref{js_def}) in a more convenient form
\begin{equation}\label{js2}
{\bf j}_s=\frac{ic^2}{4\pi\omega\lambda^2}\left({\bf E}+\nabla \tilde\Phi\right).
\end{equation}
Finally, substituting this expression to the above Maxwell equation, taking divergence of the result and combining it with imaginary part of the GL equation (\ref{GL_general}) we find 
\begin{equation}\label{phi_eq}
l_\Phi^2\nabla^2\tilde\Phi=\tilde\Phi,
\end{equation}
where $1/l_\Phi^{2}= 1/l_E^{2}-i\omega\tau/\xi^2$, $l_E=\sqrt{\hbar\sigma/\left(\pi e^2\alpha_{0}\Delta_0^2\right)}$ is the typical scale of conversion between superconducting and normal currents, $\tau=\pi\hbar/\left(8T_c\varepsilon\right)$ is the characteristic timescale of the GL theory, and $\xi=\sqrt{\hbar^2/\left(4m_{0}\alpha_{0} T_c\varepsilon\right)}$ is the superconducting correlation length. The obtained equation for the charge imbalance potential $\tilde \Phi$ does not contain sources. At the same time, the superconducting current (\ref{js2}) should not have a component perpendicular to the sample surface which means that the presence of the corresponding component in the field ${\bf E}$ of the incident wave gives rise to the nonzero gradient $\nabla\tilde\Phi$ and, thus, induces the spatial oscillation of the gap potential $\Delta$ controlled by Eq.~(\ref{GL_real}). 

The further details of the photocurrents generation depend on the sample geometry as well as on the angle of incidence and the polarization of the electromagnetic wave. For illustration, below we consider three situations, namely, (i) the photon drag effect arising in the superconducting half-space under the influence of the wave with the linear $E$ polarization, (ii) the inverse Faraday effect in superconducting disk radiated by the circularly polarized light, and (iii) photogalvanic phenomena originating from the interaction of superconductor with structured light.

\subsection*{Photon drag effect in superconducting half-space}

Here we consider a superconductor occupying the half-space $z>0$ radiated by the electromagnetic wave of the E-polarization \cite{Photon_drag}. The magnetic field outside the superconductor (for $z<0$) reads
\begin{equation}\label{wave_B_field}
B_y = B_0 e^{-i\omega t+ik \sin\theta x}\left(e^{ik\cos\theta z}+r e^{-ik\cos\theta z}\right),
\end{equation}
where $k=\omega/c$, $B_0$ is the amplitude of the incident wave, $\theta$ is the angle of incidence which we assume to be not very large, and $r$ is the complex reflection coefficient. For simplicity we restrict ourselves to the low-frequency limit assuming that (i) $\omega\ll 4\pi\sigma$ so that we may put $r\approx 1$, (ii) $\left(\omega/c\right)^2\ll {\rm min}\left\{4\pi\sigma\omega/c^2;~\lambda^{-2}\right\}$ which means that the wave-length is much larger than both the skin-layer depth and the London penetration depth, (iii) $(\omega/c)\ll\l_E^{-1}$ which corresponds to the  limit of well developed superconductivity, and (iv) $\omega/c\ll\xi^{-1}$. The above conditions also allow one to assume that inside the superconductor the magnetic field has the only nonzero component $B_y$ and, at the same time, go beyond the Leontovich boundary conditions (see, e.g., \cite{Landau_book}) by considering the component $E_z$ of electric field which induces the charge imbalance potential $\tilde \Phi$ inside the sample and gives rise to the photogalvanic phenomena. Solving Ginzburg-Landau and Maxwell equations, we find that for $z>0$
\begin{equation}\label{phis_res}
\tilde\Phi=\frac{i\omega l_\Phi B_0\sin\theta}{2\pi\sigma} \exp\left(ik\sin\theta x-z/l_\Phi\right).
\end{equation}
Then solving Eq.~(\ref{GL_real}) perturbatively with respect to the wave amplitude $B_0$, substituting the solution to the expression for the superconducting current (\ref{js2}) and, finally, integrating the resulting expression over $z$ we obtain the second order (in $B_0$) contributions to the total current density $I_x=\int\limits_0^\infty j_{sx}(z)dz$ in the form $I_{x}=I_{x}^{(0)}+I_{x}^{(\omega)}+I_{x}^{(2\omega)}$.
Here the first contribution is the dc component of the current corresponding to the photon drag effect:
\begin{equation}\label{Current_0_res_2}
I_x^{(0)}= \frac{\nu}{\sqrt{2}\pi}\frac{\xi}{\tau}\frac{B_0^2\sin\theta}{H_{cm}} \frac{\omega\tau}{\left(1+\omega^2\tau^2\eta^2\right)}{\rm Re}\left[\frac{i\sqrt{1+i\omega\tau\eta}}{\left(2-i\omega\tau\right)}\right],
\end{equation}
where $\eta=l_E^2/\xi^2$, and $H_{cm}= \Phi_0/(2\sqrt{2}\pi\xi\lambda)$ is the thermodynamic critical field. The term 
 $I_{x}^{(\omega)}$ stands for the usual linear response contribution while the contribution $I_{x}^{(2\omega)}$ describes the second harmonic response which oscillates at $2\omega$ frequency: 
\begin{equation}\label{Current_2_res_2}
I_x^{(2\omega)}= \frac{\nu}{\sqrt{2}\pi}\frac{\xi}{\tau}\frac{B_0^2\sin\theta}{H_{cm}} {\rm Re}\left[ \frac{i\omega\tau e^{2ik\sin\theta x-2i\omega t}}{\left(2-i\omega\tau\right)\left(1-i\omega\tau\eta\right)^{3/2}}\right].
\end{equation}

\rev{

Both the above current expressions strongly depend on frequency with the characteristic frequency scale $\tau^{-1}$.
For rather large frequencies $\omega\tau\gg 1$ the superconducting contribution to the rectified current becomes suppressed and  we get the crossover from the superconducting to normal metal. Note that  this crossover occurs for frequencies $\hbar\omega<2\Delta$ which can be still in the range of validity of TDGL theory.

It is interesting to address the relation between our mechanism of current rectification and the mechanisms of surface photogalvanic effect (SPGE) and photon drag effect (PDE) known in nonsuperconducting systems 
 \cite{alper,mikheev}.
 This relation 
 does not look so obvious due to the very specific physics underlying the second order nonlinear effects in superconductors. Let us consider this issue in more detail. The current in equation (\ref{Current_0_res_2}) does contain the factor $\sin\theta$ (where $\theta$ is the angle of incidence) which means that we can rewrite the current in the form $j_x\sim {k}_x |E|^2$ (the factor $\omega/c$ can be also picked out from Eq.(\ref{Current_0_res_2})). This fact seems to indicate the close relation of our effect to the PDE. On the other hand, the expression  $j_x\sim E_xE_z^*$ typical
for the SPGE current  is also relevant since the strong reflection of the electromagnetic wave from the metallic surface and resulting approximate Leontovich {boundary} condition make the electric field vector {inside the superconductor} to be almost parallel to the surface giving us again the factor $\sin\theta$ mentioned above. Considering a different polarization of the incident electromagnetic wave, i.e., taking s-polarization with the electric field vector perpendicular to the plane of incidence, we also can not distinguish clearly between the possible analogies of our effect in non-superconducting systems. Indeed, in this case our mechanism gives zero rectified current since the charge imbalance potential is not generated for this geometry. In this sense, the suppression of the effect is analogous to the suppression of the SPGE in non-superconducting systems \cite{mikheev}. On the other hand, one can see that the time averaged Hall current which underlies the PDE also vanishes in this case due to the compensation of the magnetic field components perpendicular to the surface in the incident and reflected waves. Still the analogy of our effect to the PDE looks more promising in view of the fact that both the rectified supercurrent calculated in our work and the Hall effect in superconductors are governed by the imaginary part of the relaxation constant in the time dependent Ginzburg-Landau theory. Nevertheless we cannot insist on this analogy in the absence of full microscopic analysis of the rectification mechanism. We also should note that the particular mechanism of SPGE  \cite{mikheev} which assumes the interband transitions and diffusive scattering of electrons clearly cannot be applied for the case of supercurrent rectification. To sum up, at the moment we consider the analogy with the ac Hall effect and PDE to be more relevant though we believe that only further studies on the basis of microscopic theory may help to classify the studied effect correctly.

Experimentally, the above photon drag effect can be measured, e.g., by embedding the superconducting sample into a conducting loop where the generation of the optically controlled current should be accompanied by the changes of the magnetic flux trapped inside the loop. At the same time, the second harmonic response can be detected by analyzing the spectrum of reflected electromagnetic waves.  

}

\subsection*{Inverse Faraday effect in superconducting disk}

\begin{figure}[h!]
\centerline{\includegraphics[width=0.8\columnwidth]{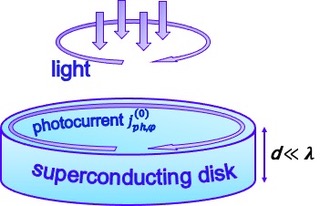}}
\caption{Illustration of the inverse Faraday effect in thin superconducting disk irradiated by the circularly polarized wave. The photoinduced circulating dc current $\mathbf{j}_{ph, \varphi}^{(0)}$ creates the magnetic moment of the disk.}
\label{disk}
\end{figure}

Somewhat similar situation is realized in a thin superconducting disc of the radius $R$ and thickness $d$ (considered to be much less than the London penetration depth $\lambda$) irradiated by the circularly polarized wave which propagates perpendicular to the disk surface (see Fig.~\ref{disk}) and has the electric field acting on electrons inside the superconductor ${\bf E}=E_0 \ {\rm Re}\left[({\bf e}_x+i{\bf e}_y)e^{-i\omega t}\right]$. Here we choose the origin of the coordinate system with the in-plane axes $x$ and $y$ in the disk center) \cite{IFE1,IFE2}. Although the electric field is parallel to the disk surface, obviously, it has a nonzero projection normal to the  disk edge. According to the above analysis, this gives rise to the charge imbalance potential and, therefore, to the generation of the circulating dc current. Further calculations of the charge imbalance potential are based on the equation (\ref{phi_eq}) and, thus, we generalize here the results of Ref.\cite{IFE1} for arbitrary ratios $l_E/R$.

\begin{figure}[hbt!]
\centerline{\includegraphics[width=1.0\columnwidth]{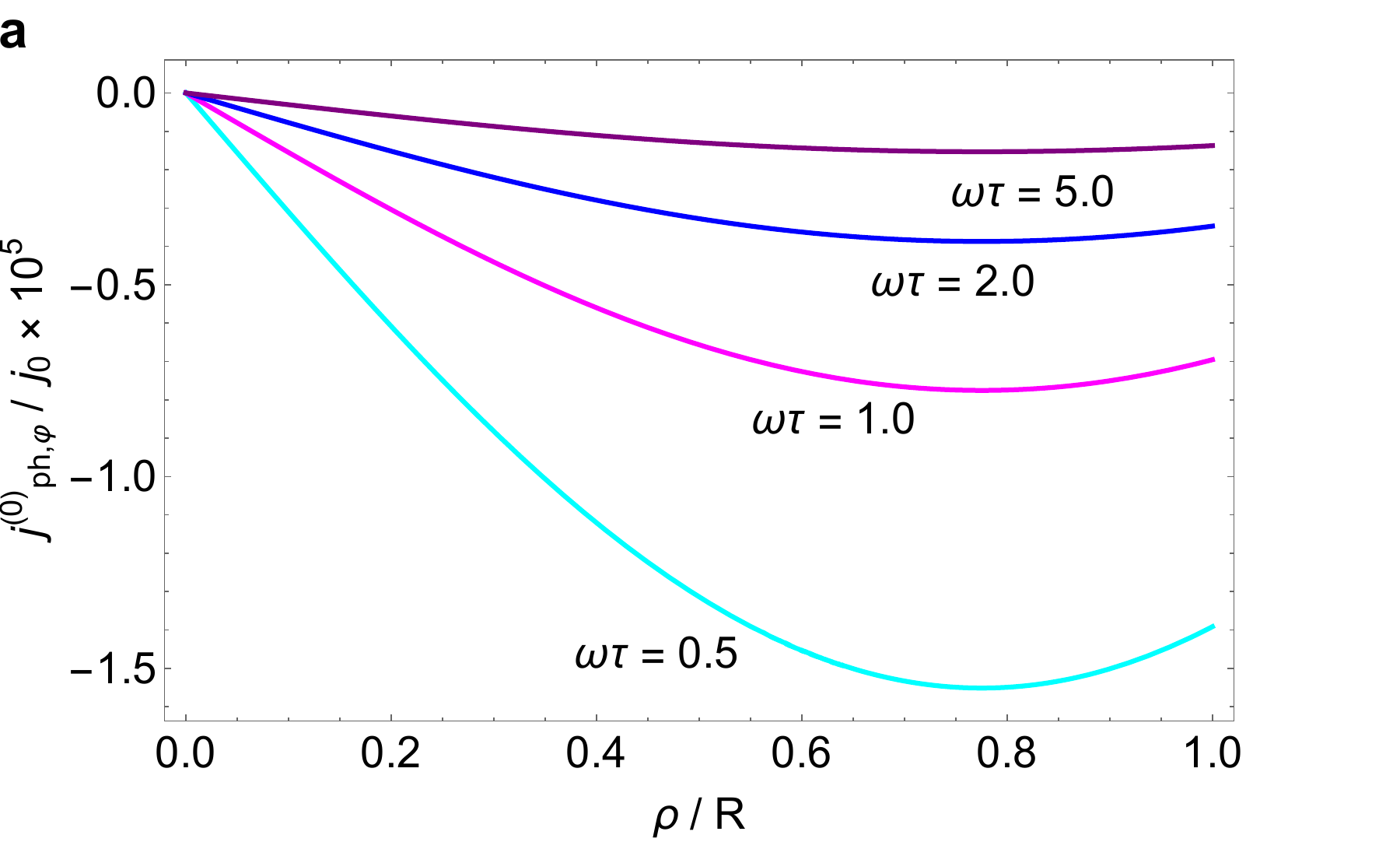}}
\hfill
\centerline{\includegraphics[width=1.0\columnwidth]{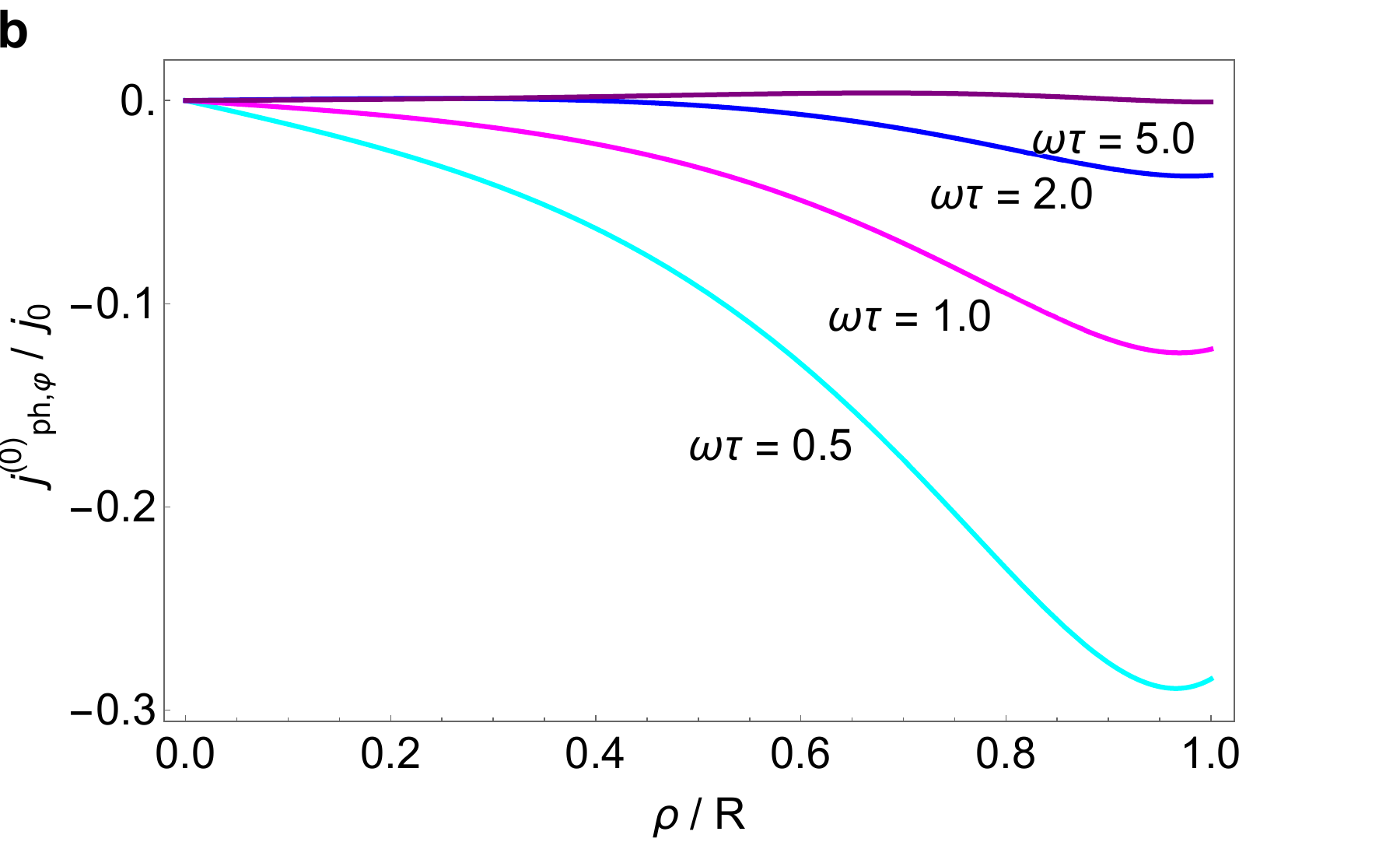}}
\caption{Dependence of azimuthal dc photocurrent on the distance from the disk center for $R=0.1\xi$ (top) and $R=3\xi$ (bottom), $l_E/\xi = 1/\sqrt{5.79}$ and different values of the normalized frequency $\omega \tau$.}
\label{disk_cur}
\end{figure}

Solving the Ginzburg-Landau equation together with the boundary condition at the edge of the disk $(\partial_\rho\tilde{\Phi} + E_\rho)|_{\rho = R} = 0$ (here we use the polar coordinates $\rho$ and $\varphi$), we find that $\tilde{\Phi} = -E_0 R f\left(q_2, \rho\right)$, where
\begin{equation}\label{f_def}
f\left(q, \rho\right) = \frac{J_1\left(q\rho\right)}{qR J_0\left(q R\right) - J_1\left(q R\right)},
\end{equation}
$q_2^2 = -1/l_\Phi^2$ [we choose ${\rm Im}(q_2)>0$], $J_0$ and $J_1$ are Bessel functions. Then solving Eq.~(\ref{GL_real}) perturbatively with respect to the wave amplitude $E_0$, substituting the solution to the expression for the superconducting current (\ref{js2}) and considering the contribution at zero frequency, we obtain the expression for azimuthal dc photoinduced current: 
\begin{equation}
\begin{array}{c}{\displaystyle j_{ph,\varphi}^{(0)} = j_0 \frac{1}{\omega \tau}\frac{R}{\xi} \frac{2l_E^2}{2l_E^2 - \xi^2}}\\{}\\{\displaystyle \times {\rm Re}\bigg\{ \bigg[f\left(q_1, \rho\right) - f\left(q_2, \rho\right)\bigg]\left[1 - \frac{R}{\rho} f\left(q_2^*, \rho\right)\right] \bigg\} ,}
\end{array}
\end{equation}
where $j_0 =  {4 e^3 \Delta_0^2 E_0^2 \tau^3\nu / \pi \alpha_0\xi m_0^2  }$ and $q_1^2 = i\omega\tau/\xi^2 - 2/\xi^2$ [we choose ${\rm Im} \left(q_1\right) > 0$]. Typical current density profiles for different disk radii are shown in Fig.~\ref{disk_cur}.

The circulating dc current, obviously, creates a dc magnetic moment $M_{ph} = \left(d/c\right) \int_{0}^{R} j_{ph,\varphi}^{(0)} \pi \rho^2 d\rho$. The generation of such magnetic moment by circularly polarized electromagnetic wave is, in fact, a manifestation of the inverse Faraday effect. In the limit of small disks ($R \ll \xi$) one finds:
\begin{equation}
        \label{momsmallas}
 M_{ph} = M_0 \left\{
    \begin{aligned}
    & -\frac{11}{256} \left (\frac{R}{l_{E}}\right) ^2 \left (\frac{R}{\xi} \right )^2 \frac{1}{\omega \tau}, & \omega \tau \ll \left (\frac{\xi}{R} \right ) ^2\\
   &  \frac{2}{(\omega \tau)^3} \left (\frac{\xi}{R}\right )^2, & \omega \tau \gg \left ( \frac{\xi}{R} \right )^2
    \end{aligned}
     \right. ,
\end{equation}
where $M_0 = \left(j_0 R{d}/c\right) \pi R^2 \left(R/\xi\right)$. This photoinduced magnetic moment will be, of course, partially screened by the dc Meissner currents. We omit this effect assuming the disk radius $R$ to be much less than the effective screening length $\lambda^2 / d$.  

In addition, one can find the radial dc photocurrent $j_{ph, \rho}^{(0)}$ in a full analogy with the calculation of the azimuthal one. This current has non-zero divergence, and it should be compensated by another current in order to prevent the time-dependent charge accumulation. In superconductors, the compensating mechanism arises from the inhomogeneous distribution of the zero-frequency superconducting phase $\chi^{(0)}(\rho)$. In case of a thin superconducting disk this phase can be obtained from the equation 
\begin{equation}
\frac{4 \pi \lambda^2}{c}j_{ph, \rho}^{(0)} + \frac{\hbar c}{2 e} \nabla \chi^{(0)}= 0.
\end{equation}
Thus, the superconducting phase difference arising between the center and the edge of the disk opens the way for using such system as a phase battery \cite{samokh,aronov}. 

\subsection*{Interaction of superconductor with structured light}

Another interesting situation occurs in superconductors irradiated with the so-called twisted light, which can be realized in Bessel or Laguerre-Gaussian beams \cite{allen1,forbes,allen2,molina,Knyazev}. The key feature of twisted light is its non-zero orbital momentum which can be transfered to the electrons in conductive media \cite{taras3}. We will consider the interaction of the Bessel beam characterized by an angular momentum ${L}$ with the bulk superconductor occupying the half-space $z > 0$. Below we show that the induced photocurrent contains contributions from both the inverse Faraday effect due to the polarization rotation and the photon drag effect due to the transfer of orbital momentum.  


Let us remind that the Bessel beam is defined as a superposition of plane waves with wave vectors $\bf k$ lying on a cone with a fixed angle $\vartheta$ to the propagation axis $z$. The magnetic field of the Bessel beam in vacuum can be written as follows \cite{allen1,forbes,allen2,molina,Knyazev,Oleg}: 

\begin{multline}
\label{HApp}
\mathbf{H}^{(e)} =
\begin{pmatrix}
H_\rho^{(e)} \\
H_\varphi^{(e)} \\
H_z^{(e)}
\end{pmatrix}
=
\frac{B_0}{2} e^{-i\omega t + ik_z z + i{L} \varphi} \times \\
\times \begin{pmatrix}
- J_{{L}+1}(\kappa \rho) o_- - J_{{L}-1}(\kappa \rho) o_+ \\
i J_{{L}+1}(\kappa \rho) o_- - i J_{{L}-1}(\kappa \rho) o_+ \\
-2b \sin\vartheta J_{L}(\kappa \rho) 
\end{pmatrix},
\end{multline}
where $\kappa = k \sin \vartheta$, $o_\pm = a \pm i b \cos\vartheta$ and $a$, $b$ are the components of a mixed polarization unit vector of plane waves forming a beam, corresponding to the contributions of {p-} and {s-}polarizations.

Using Maxwell equations inside the bulk superconductor in the low-frequency limit $\left(\omega/c\right)^2\ll {\rm min}\left\{4\pi\sigma\omega/c^2;~\lambda^{-2}\right\}$, taking into account the reflected wave (assuming reflection coefficients for {p-} and {s-}polarizations $r_\parallel, r_\perp \approx 1$) and continuity of the tangential component of the magnetic field at the boundary $z = 0$, we can neglect the z-component of the magnetic field and obtain the field inside the superconducting half-space:
\begin{equation}
\label{hbeamin}
\mathbf{H}^{(i)}=
B_0 e^{-i\omega t + i{L} \varphi} e^{-z/\lambda_\text{eff}}
\begin{pmatrix}
-J_{{L}+1}(\kappa \rho) o_- - J_{{L}-1}(\kappa \rho) o_+ \\
iJ_{{L}+1}(\kappa \rho) o_- - iJ_{{L}-1}(\kappa \rho) o_+ \\
0
\end{pmatrix},
\end{equation}
where $\lambda_\text{eff}^{-2} = \lambda^{-2} \left(1-i\omega\tau l_E^2/\xi^2\right)$.

Next, by analogy with the previous sections, one can obtain the charge imbalance potential $\tilde{\Phi}$, the gap function $\Delta$ and, therefore, the dc photocurrent which contains azimuthal, radial and vertical components. For example, the expression for the azimuthal dc photocurrent takes the form:
\begin{equation}
\label{currentphzero}
\begin{array}{c}{\displaystyle
\small j^{(0)}_{\text{ph},\varphi} =\small \frac{\nu B_0^2 \omega\sin\vartheta}{\sqrt{2} \pi H_{cm}} \frac{\xi}{\lambda \ (\xi^2-2 l_{E}^2)}
{\rm Re} \Bigg\{i \lambda_{\rm eff}^* \bigg[ \frac{e^{iq_1 z}}{q_1} - \frac{e^{iq_2 z}}{q_2} \bigg]}\\{}\\{\displaystyle  \times \left[ab^* \cos\vartheta \frac{J_{L}(\kappa\rho)}{\kappa}\frac{\partial J_{L}(\kappa \rho)}{\partial\rho} + \frac{i{L}|a|^2}{\kappa \rho}J^2_{{L}}(\kappa \rho)\right] \Bigg\},}
\end{array}
\end{equation}
where the expressions for the values $q_1$ and $q_2$ are the same as in the previous subsection. 

One sees that the expression (\ref{currentphzero}) contains two contributions. The term proportional to $\frac{J_{L}(\kappa\rho)}{\kappa}\frac{\partial J_{L}(\kappa \rho)}{\partial\rho}$ requires the presence of both {p-} and {s-}polarizations and can be viewed as a manifestation of the inverse Faraday effect. The term proportional to $\left({L}/\kappa \rho\right)J^2_{{L}}\left(\kappa \rho\right)$ is antisymmetric under the transformation ${L} \to -{L}$ and originates from the photon drag effect.

It should be mentioned that since the generation of the charge imbalance potential $\tilde{\Phi}$ requires the presence of an electric field component perpendicular to the sample surface, the plane waves forming the beam must have a {p-}polarization contribution.

The photoinduced dc current is screened by the Meissner current $\mathbf{j}^{(0)}_\text{M}$ providing the cancellation of the dc magnetic field $\mathbf{B}^{(0)}$ inside the superconductor at the distance of the London penetration depth $\lambda$ from the surface $z = 0$. Besides that, as was mentioned above for the case of a disk, since dc photocurrents have non-zero divergence it should be compensated by an inhomogeneous superconducting phase $\chi^{(0)}$. All these dc currents and magnetic field can be obtained from the electrodynamic system of equations:
\begin{equation}
\label{Meisnersistem2}
\left\{\begin{array}{l}{
\displaystyle {\rm curl~} {\rm curl~}\mathbf{A}^{(0)} = \frac{4\pi}{c}\left(\mathbf{j}_\text{M}^{(0)}+\mathbf{j}_\text{ph}^{(0)}\right),
}\\{}\\{  \displaystyle
\frac{4\pi \lambda^2}{c} \mathbf{j}_\text{M}^{(0)}=- \mathbf{A}^{(0)} +\frac{\hbar c}{2e} \nabla\chi^{(0)}\ ,} 
\end{array}\right.
\end{equation}
where ${\rm curl~}\mathbf{A}^{(0)} = \mathbf{B}^{(0)}$.

\begin{figure}[hbt!]
\centerline{\includegraphics[width=1.0\columnwidth]{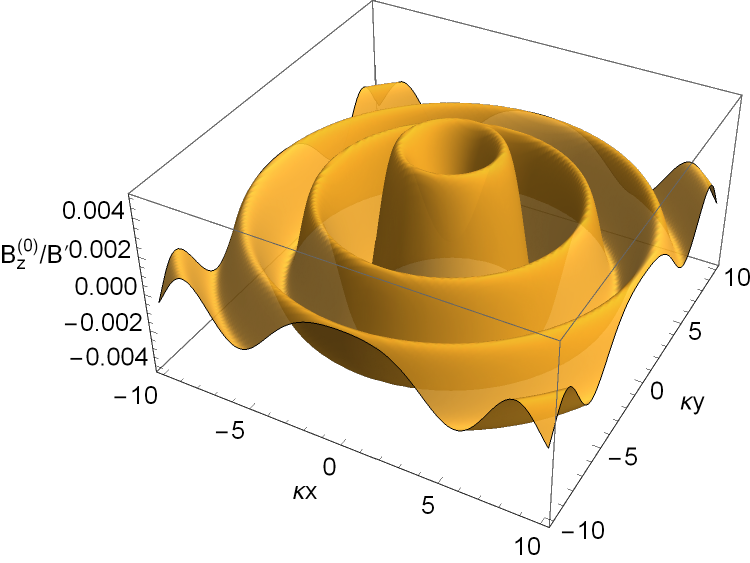}}
\caption{Magnetic field profiles $B_z(x,y)$. Here we put $\lambda/\xi = 2$, $l_E/\xi = 1/\sqrt{5.79}$, $\omega\tau=1$,  $B'=4\pi\nu B_0^2/\sqrt{2} H_\text{cm}$. The values $a=1/\sqrt{2}, b=i/\sqrt{2}, l=2$ are used.}
\label{Hzprof}
\end{figure}

Detailed calculations performed in a recent paper \cite{Oleg} show that the presence of the screening Meissner current leads to the total compensation of the magnetic moment created by dc photocurrent, so that the resulting magnetic moment is  zero. At the same time, the components of the dc magnetic field $\mathbf{B}^{(0)}$ are non zero and can be measured experimentally. Typical spatial profiles of the $z$-component of the dc magnetic field are shown in Fig.~\ref{Hzprof}.

\section*{Conclusion}

To sum up, we have presented a review of our recent works on the phenomenological description of the
mechanisms of second order nonlinear response of superconductors irradiated by electromagnetic waves in the microwave and THz frequency range. The resulting rectification effect for electromagnetic radiation of different polarization and orbital momenta is shown to give rise to dc photocurrents, magnetic moments and magnetic fields, as well as provides the possibility to realize the setup with the dc superconducting phase differences tuned by electromagnetic radiation. This photoinduced phase battery effect
can be exploited, e.g., in multiply connected geometries for switching between the superconducting states with different vorticities. 
The effect of photoinduced switching between vortex states according to the above mechanisms provides a convenient tool for superconducting optofluxonics allowing one to manipulate the magnetic flux trapped inside the loop which is promising for applications in the devices of the rapid single flux quantum (RSFQ) logics.

\section*{Acknowledgements}
This work was supported by the Russian Science Foundation (Grant No. 25-12-00042) in part of the analysis of the effects caused by the charge imbalance potential, by the Grant of the Ministry of science and higher education of the Russian Federation No. 075-15-2025-010 in part of the study of superconductors with intrinsic diode effect, and by MIPT Project No. FSMG-2023-0014 in part related to the interaction of superconductors with structured light. S. V. M. and M. V. K. also acknowledge the financial support of the Foundation for the Advancement of Theoretical Physics and Mathematics BASIS (Grant No. 23-1-2-32-1).

\section*{Information regarding all authors}

In the order of appearance:

\smallskip
S.~V.\ Mironov, Institute for Physics of Microstructures, Russian Academy of Sciences, 603950 Nizhny Novgorod, GSP-105, Russia, 0000-0001-5378-6105

\smallskip
A.~I.\ Buzdin, University Bordeaux, LOMA UMR-CNRS 5798, F-33405 Talence Cedex, France, Institute for Computer Science and Mathematical Modeling, Sechenov First Moscow State Medical University, 119991 Moscow, Russia, 0000-0001-7615-8785

\smallskip
O.~B.\ Zuev, Moscow Institute of Physics and Technology (National research university), Dolgoprudny, Moscow region 141700, Russia, L.~D.\ Landau Institute for Theoretical Physics,  Chernogolovka 142432, Russia, 0009-0007-2936-9812

\smallskip

M.~V.\ Kovalenko, Moscow Institute of Physics and Technology (National research university), Dolgoprudny, Moscow region 141700, Russia, L.~D.\ Landau Institute for Theoretical Physics,  Chernogolovka 142432, Russia, 0009-0003-9518-2707

\smallskip
A.~S.\ Mel'nikov, Moscow Institute of Physics and Technology (National research university), Dolgoprudny, Moscow region 141700, Russia, Institute for Physics of Microstructures, Russian Academy of Sciences, 603950 Nizhny Novgorod, GSP-105, Russia, 0000-0002-4241-467X

\section*{Corresponding author}

S.~V.\ Mironov, svmironov@ipmras.ru


\begin{thebibliography}{99}

\bibitem{IFE1} Mironov S. V., Mel'nikov A. S., Tokman I. D., Vadimov V., Lounis B., Buzdin A. I. {\sl Inverse Faraday effect for superconducting condensates} Phys. Rev. Lett., vol. 126, 137002 (2021); DOI: 10.1103/PhysRevLett.126.137002.

\bibitem{plast} Plastovets V. D., Tokman I. D., Lounis B.,
Mel'nikov A. S., and Buzdin A. I. {\sl All-optical generation
of Abrikosov vortices by the inverse Faraday
effect} Phys. Rev. B vol. 106, 174504 (2022); DOI: 10.1103/PhysRevB.106.174504.


\bibitem{croit} Croitoru M. D., Mironov S. V., Lounis B., and
Buzdin A. I. {\sl Toward the light-operated superconducting devices:
Circularly polarized radiation manipulates the
current-carrying states in superconducting rings} Adv.
Quantum Technol. vol. 5, 2200054 (2022); DOI: 10.1002/qute.202200054.

\bibitem{croit2} Croitoru M. D., Lounis B., Buzdin A. I., {\sl Helicity-controlled switching of superconducting states by radiation pulse} Appl. Phys. Lett., vol. 123, 122601 (2023); DOI: 10.1063/5.0165874

\bibitem{IFE2} Putilov A. V., Mironov S. V., Mel'nikov A. S., Bespalov A. A. {\sl Inverse Faraday Effect in Superconductors with a Finite Gap in the Excitation Spectrum} JETP Letters, vol. 117, 827 (2023); DOI: 10.1134/S0021364023601239.

\bibitem{Photon_drag} Mironov S. V., Mel'nikov A. S., Buzdin A. I. {\sl ac Hall effect and photon drag of superconducting condensates} Phys. Rev. Lett., vol. 132, 096001 (2024); DOI: 10.1103/PhysRevLett.132.096001.

\bibitem{mironov-diode} Mironov S. V., Mel'nikov A. S., Buzdin A. I. {\sl Photogalvanic phenomena in superconductors supporting intrinsic diode effect} Phys. Rev. B, vol. 109, L220503 (2024); DOI: 10.1103/PhysRevB.109.L220503.
    
\bibitem{Oleg} Zuev O. B., Kovalenko M. V., Mel'nikov A. S. {\sl Superconducting photocurrents induced by structured electromagnetic radiation} Phys. Rev. B, vol. 112, 144503 (2025); DOI: 10.1103/cr1g-wnmc.    
    
    
    

\bibitem{Belinicher_rect} Belinicher V. I., Sturman B. I. {\sl The photogalvanic effect in media lacking a center of symmetry} Sov. Phys. Usp., vol. 23, 199 (1980); DOI: 10.1070/PU1980v023n03ABEH004703.

\bibitem{Ivchenko_rect} Ivchenko E., Ganichev  S., {\sl in Spin Physics in Semiconductors} edited by M. I. Dyakonov (Springer, Berlin, 2008), Chap. 9.

\bibitem{Glazov_rect} Glazov M. M., Ganichev S. D. {\sl High frequency electric field induced nonlinear effects in graphene} Phys. Rep., vol. 535, 101 (2014); DOI: 10.1016/j.physrep.2013.10.003.

\bibitem{tokura} Tokura Y. and Nagaosa N. {\sl Nonreciprocal responses from noncentrosymmetric
quantum materials} Nat. Commun. vol. 9, 3740 (2018); DOI: 10.1038/s41467-018-05759-4.


\bibitem{taras} Gunyaga A. A., Durnev M. V., and Tarasenko S. A. {\sl  Second
harmonic generation due to the spatial structure of
a radiation beam} Phys. Rev. Lett. vol. 134, 156901 (2025); DOI: 10.1103/PhysRevLett.134.156901.    

\bibitem{perel} Perel V. I., Pinsky Ya. M. {\sl  Optical orientation of electrons
in semiconductors} Sov. Phys. Solid State, vol. 15, 996 (1973).

\bibitem{barlow}  Barlow H. M. {\sl  The Hall effect and its application to microwavepower measurement} Proc. IRE  vol. 46, 1411 (1958); DOI: 10.1109/JRPROC.1958.287010.

\bibitem{normantas} Normantas E. and Pikus G. E. {\sl Drag effect at light reflection from the surface}  Fiz. Tverd. Tela 27, 3017 (1985) [Sov. Phys. Solid State 27, 1762 (1985)].

\bibitem{taras2} Durnev M. V., Tarasenko S. A. {\sl Rectification of AC electric current at the edge of a two-dimensional electron gas}  Phys. Status Solidi B, vol. 258, 2000291 (2021); DOI: 10.1002/pssb.202000291.


\bibitem{gurevich1} Gurevich V. L., Laiho R. {\sl Photomagnetism of metals: Microscopic theory of the photoinduced surface current} Phys. Rev. B, vol. 48, 8307 (1993); DOI: 10.1103/PhysRevB.48.8307.


\bibitem{gurevich2} Gurevich V. L., Thellung A. {\sl Photomagnetism of metals: Microscopic theory of photoinduced bulk current}  Phys. Rev. B, vol. 49, 10081 (1994); DOI: 10.1103/PhysRevB.49.10081.


\bibitem{ivchenko} Ivchenko E. L. {\sl Photoinduced currents in graphene and carbon nanotubes}  Phys. Status Solidi B, vol. 249, 2538 (2012); DOI: 10.1002/pssb.201200081.


\bibitem{taras3} Gunyaga A. A., Durnev M. V., and Tarasenko S. A. {\sl Photocurrents
induced by structured light}  Phys. Rev. B, vol. 108, 115402 (2023); DOI: 10.1103/PhysRevB.108.115402.

\bibitem{yeh1} T.-T. Yeh, H. Yerzhakov, L. Bishop-Van Horn, S. Raghu, and A. Balatsky {\sl Structured light and induced vorticity in superconductors I: Linearly polarized light} arXiv:2407.15834 (2024); DOI: 10.48550/arXiv.2407.15834.


\bibitem{yeh2} T.-T. Yeh, H. Yerzhakov, L. Bishop-Van Horn, S. Raghu, and A. Balatsky {\sl Structured light and induced vorticity in superconductors II: Quantum print with Laguerre-Gaussian beam}  arXiv:2412.00935 (2024);  DOI: 10.48550/arXiv.2412.00935.


\bibitem{Likharev_Semenov} Likharev K. K., Semenov V. K. {\sl RSFQ logic/memory family: a new Josephson-junction technology for sub-terahertz-clock-frequency digital systems} IEEE Trans. Appl. Supercond., vol. 1, 3 (1991); DOI: 10.1109/77.80745.

\bibitem{Krasnov_DE} Krasnov V. M., Oboznov  V. A., Pedersen  N. F. {\sl Fluxon dynamics in long Josephson junctions in the presence of a temperature gradient or spatial nonuniformity} Phys. Rev. B, vol. 55, 14486 (1997); DOI: 10.1103/PhysRevB.55.14486.

\bibitem{Villegas_DE} Villegas J. E., Savel'ev S., Nori F., Gonzalez E. M., Anguita J. V., Garcia R., Vicent J. L. {\sl A superconducting reversible rectifier that controls the motion of magnetic flux quanta} Science, vol. 302, 1188 (2003); DOI: 10.1126/science.1090390.

\bibitem{ando} F. Ando, Y. Miyasaka, T. Li, J. Ishizuka, T. Arakawa, Y. Shiota,
T. Moriyama, Y. Yanase, and T. Ono {\sl Observation of superconducting
diode effect} Nature (London), vol. 584, 373 (2020); DOI: 10.1038/s41586-020-2590-4.

\bibitem{Nadeem_rev} Nadeem M., Fuhrer M. S., Wang X. {\sl The superconducting diode effect} Nat. Rev. Phys., vol. 5, 558 (2023); DOI: 10.1038/s42254-023-00632-w.
    
    
\bibitem{daido}  Daido A., Ikeda Y., and Yanase Y.  {\sl Intrinsic superconducting
diode effect} Phys. Rev. Lett., vol. 128, 037001 (2022); DOI: 10.1103/PhysRevLett.128.037001.

\bibitem{he}  He J. J., Tanaka Y., and Nagaosa N. {\sl  A phenomenological
theory of superconductor diodes} New J. Phys., vol. 24, 053014 (2022); DOI: 10.1088/1367-2630/ac6766.

\bibitem{edel1}  Edelstein V. M. {\sl  The Ginzburg-Landau equation for superconductors of polar symmetry} J. Phys.: Condens. Matter, vol. 8, 339 (1996); DOI: 10.1088/0953-8984/8/3/012.

\bibitem{edel2}  Edelstein V. M. {\sl  Magnetoelectric effect in polar superconductors} Phys. Rev. Lett., vol. 75, 2004 (1995); DOI: 10.1103/PhysRevLett.75.2004.

\bibitem{persh}  Pershoguba S. S., Bj\"{o}rnson K., Black-Schaffer A. M., and Balatsky A. V. {\sl Currents induced by magnetic impurities in superconductors with spin-orbit coupling} Phys. Rev. Lett., vol. 115, 116602 (2015); DOI: 10.1103/PhysRevLett.115.116602.

\bibitem{mironov-B}  Mironov S. and Buzdin A.  {\sl Spontaneous currents in superconducting systems with strong spin-orbit coupling} Phys. Rev. Lett., vol. 118, 077001 (2017); DOI: 10.1103/PhysRevLett.118.077001.   

\bibitem{Pitaevskii61} Pitaevskii L. P. {\sl Electric Forces in a Transparent Dispersive Medium} JETP, vol. 12, 1008 (1961).

\bibitem{boev1} Boev M. V. {\sl Photon drag of superconducting fluctuations in two-dimensional systems} Phys. Rev. B, vol. 101, 104512 (2020); DOI: 10.1103/PhysRevB.101.104512.

\bibitem{boev2} Parafilo A. V., Boev M. V., Kovalev V. M., and Savenko I. G. {\sl Photogalvanic transport in fluctuating Ising superconductors} Phys. Rev. B, vol. 106, 144502 (2022); DOI: 10.1103/PhysRevB.106.144502.

\bibitem{boev3} Boev M. V. and Kovalev V. M. {\sl Contribution of fluctuations of the order parameter to the second harmonic generation in two-dimensional monomolecular superconductors}
Pis'ma Zh. Eksp. Teor. Fiz., vol. 116, 173 (2022) [JETP Lett., vol. 116, 173 (2022)] ; DOI: 10.1134/S0021364022601269.

\bibitem{Sonowal} Sonowal K., Parafilo A. V., Boev M. V., Kovalev V. M., Savenko I. G. {\sl Second-harmonic generation in fluctuating Ising superconductors} 2D Mater., vol. 10, 045004 (2023); DOI: 10.1088/2053-1583/ace45c.

\bibitem{Parafilo_2025} Parafilo A. V., Kovalev V. M., Savenko I. G. {\sl Ratchet Hall Effect in Fluctuating Superconductors} (2025); DOI: 10.48550/arXiv.2505.04478

\bibitem{Plastovets2} \rev{Plastovets V., Buzdin A. {\sl Magnetoelectric effect in the helical state of a superconductor/ferromagnet bilayer} Phys. Rev. B, vol. 110, 144521 (2024); DOI: 10.1103/PhysRevB.110.144521.}

\bibitem{Levchenko} \rev{Levchenko A., Hasan J., Shaffer D., Khodas M. {\sl Supercurrent diode effect in helical superconductors} Phys. Rev. B, vol. 110, 024508 (2024); DOI: 
10.1103/PhysRevB.110.024508.}

\bibitem{samokh} Robinson J. W. A.,  Samokhvalov A. V., and Buzdin A. I. {\sl  Chirality-controlled spontaneous currents in spin-orbit coupled superconducting rings} Phys. Rev. B, vol. 99, 180501(R) (2019); DOI: 10.1103/PhysRevB.99.180501.

\bibitem{aronov} Aronov A. G., Gal'perin Yu. M., Gurevich V. L., Kozub V. I. {\sl  Nonequilibrium Superconductivity} (Elsevier, Amsterdam, 1986).

\bibitem{Radkevich} Radkevich A. A., Semenov A. G. {\sl Nonlinear microwave response of clean superconducting films} Phys. Rev. B, vol. 106,
094505 (2022); DOI: 10.1103/PhysRevB.106.094505.

\bibitem{plast2} Plastovets V.,  Buzdin A. I., {\sl Fluctuation-mediated inverse Faraday effect in superconducting rings} Phys. Lett. A, vol. 481, 129001 (2023); DOI: 10.1016/j.physleta.2023.129001.

\bibitem{Landau_book} Landau L. D., Lifshitz E. M. {\sl Electrodynamics of
Continuous Media} 2nd ed. (Pergamon, New York, 1984), Vol. 8.


\bibitem{alper} \rev{Alperovich V. L., Belinicher V. I., Novikov V. N., Terekhov A. S. {\sl Photogalvanic effects investigation in gallium arsenide}, Ferroelectrics, vol. 45, 1 (1982); DOI: 10.1080/00150198208208275.}

\bibitem{mikheev} \rev{Mikheev G. M., Saushin A. S., Styapshin V. M., Svirko Y. P. {\sl Interplay of the photon drag and the surface photogalvanic effects in the metal-semiconductor nanocomposite}, Scientific Reports, vol. 8, 8644 (2018); DOI: 10.1038/s41598-018-26923-2.}


\bibitem{allen1} Allen L., Beijersbergen M. W., Spreeuw R. J. C., Woerdman  J. P. {\sl Orbital angular momentum of light and the transformation of Laguerre-Gaussian laser modes} Phys. Rev. A, vol. 45, 8185 (1992); DOI: 10.1103/physreva.45.8185.

\bibitem{forbes} Forbes A., de Oliveira M., Dennis M. R. {\sl Structured
light} Nat. Photonics, vol. 15, 253 (2021); DOI: 10.1038/s41566-021-00780-4.

\bibitem{allen2} Allen L., Padgett M. J., Babiker M. {\sl The orbital angular momentum of light} in Progress in Optics, edited by E. Wolf (Elsevier, Amsterdam, 1999), vol. 39, 291; DOI: 10.1016/S0079-6638(08)70391-3.

\bibitem{molina} Molina-Terriza G., Torres J. P., Torner L. {\sl Twisted
photons} Nat. Phys., vol. 3, 305 (2007); DOI: 10.1038/nphys607.


\bibitem{Knyazev} Knyazev B. A., Serbo V. G. {\sl Beams of photons with nonzero orbital angular momentum projection: new results} UFN, vol. 188, pp. 508-539 (2018); DOI:10.3367/UFNr.2018.02.038306.








\end{thebibliography}
\end{document}